\documentclass[prl,twocolumn,reprint,elide,hyperlink]{revtex4-1}
\pdfoutput=1
\usepackage{graphicx}
\usepackage{amsmath}
\usepackage[varg]{txfonts}
\usepackage[usenames]{color}

\newcommand{\units}[1]{\,\mathrm{#1}}

\newcommand{\dexp}[1]{\cdot 10^{#1}}

\definecolor{DarkRed}{rgb}{0.6,0,0}

\newcommand{\laplacian}{\nabla^2}
\newcommand{\mysection}[1]{\textit{#1}.-}

\newcommand{\iaa}{
IAA-CSIC, P.O. Box 3004, 18080 Granada, Spain}
\newcommand{\cwi}{
CWI, P.O. Box 94079, 1090 GB Amsterdam, The Netherlands}
\newcommand{\eindhoven}{
Dept. Physics, Eindhoven Univ. Techn., The Netherlands}

\begin{document}

\title{Electron density fluctuations accelerate the branching of streamer discharges in air}
\author{A. Luque$^1$,\
U. Ebert$^{2,3}$\
}
\affiliation{$^1$\iaa,\
\\$^2$\cwi,\
\\$^3$\eindhoven\
}

\date{\today}

\begin{abstract}
Branching is an essential element of streamer discharge dynamics 
but today it is understood only qualitatively.  The variability and 
irregularity observed in branched streamer trees suggest that
stochastic terms are relevant for the description of streamer branching.
We here consider electron density fluctuations due to the discrete
particle number as a source of stochasticity in positive streamers in air at 
standard temperature and pressure. We derive a quantitative estimate
for the branching distance that agrees within a factor of 2 with 
experimental values.  As branching without noise would occur later, 
if at all, we conclude that stochastic particle noise is relevant for 
streamer branching in air at atmospheric pressure.
\end{abstract}

\maketitle
 Streamers are filamentary electrical discharges that propagate through a 
non-conducting medium when it is suddenly exposed to a high electric
field \cite{Raether1939/ZPhy}.  As first stages of electrical
breakdown \cite{Raizer1991/book}, they are found in nature preceding a
lightning stroke and as the building blocks of
upper-atmospheric discharges above thunderstorms 
\cite{Franz1990/Sci, Pasko2003/Natur, *Pasko1998/GeoRL/1, Ebert2010/JGRA/c}.
Due to their efficient production of chemical radicals \cite{Winands2006/JPhD},
streamers are used in industry for gas cleaning and sterilization.

Streamer discharges frequently form irregular trees with many branches
both in laboratory, where
branch length and branching angles in air have been measured recently 
\cite{Briels2008/JPhD, Nijdam2008/ApPhL}, and
in upper-atmospheric discharges \cite{McHarg2010/JGRA/c}.
However, our present understanding of streamer branching is only
qualitative.  Streamer trees in the laboratory are highly random and
irregular but simulations have shown that streamers can branch even in
fully deterministic density models \cite{Arrayas2002/PhRvL,
  Montijn2006/PhRvE, Liu2006/JPhD, Babaeva2008/ITPS}.  Model
reduction and analytical theory explained branching as a Laplacian 
instability that develops when the space charge layer around the
streamer head is much thinner than the streamer radius
\cite{Ebert2011/Nonli}. Nevertheless, 
the analysis of a reduced moving-boundary streamer model suggests that streamer
heads are linearly stable \cite{Tanveer2009/PhyD, Kao2010/PhyD} even if the
stabilizing effects of electron diffusion and photo-ionization are
neglected.  Streamer fingers are
therefore similar to laminar pipe flow: a finite perturbation is
required to let streamers branch or to make the pipe flow
turbulent. 

A small, but finite perturbation to trigger branching can be due to
the initial condition. But it can also be created by fluctuations
during the evolution. Such fluctuations are naturally included in
Monte Carlo models \cite{Moss2006/JGRA, Chanrion2008/JCoPh, 
*Chanrion2010/JGRA/c,  Li2009/JPhD, *Li2010/JCoPh, *Li2011/arXiv} that
follow the single electron motion; they model the stochastic
distribution of electron positions and energies. Those models
are microscopically very accurate at the cost of high computational
demands.  Even state-of-the art spatially hybrid codes
\cite{Li2009/JPhD, *Li2010/JCoPh, *Li2011/arXiv} are still limited to 
quite short streamers.

Here we introduce a new computational model and present its quantitative
predictions for positive streamers in air at standard
temperature and pressure. The model is a spatially extended stochastic
model \cite{Gardiner2004/book} that assumes an electron energy
distribution determined by the local electric field but accounts
for density fluctuations due to the discrete number of
electrons.   Since particles are not individually tracked, the
required memory and computations are roughly independent of the number of
active particles, in contrast to Monte Carlo methods.  We study the
evolution towards branching with realistic density
fluctuations in full three dimensions and, extrapolating, obtain branching
ratios that are consistent with the experiments
\cite{Briels2008/JPhD, Nijdam2008/ApPhL}.  This suggests that 
streamer branching in air at atmospheric pressure is triggered 
by electron density fluctuations. 

\mysection{Model}
  The most relevant microscopic processes in streamers are two-body
reactions between free electrons and neutral gas molecules.  Therefore
most quantities of a streamer discharge scale with the neutral gas
density in a definite manner called Townsend scaling \cite{Ebert2010/JGRA/c}.
Following \cite{Ebert2006/PSST} we define a typical 
length for streamers in air $l_0 \approx 2.3 \units{\mu m} \cdot (N_0/N)$,
a typical electric field $E_0 \approx 
2\dexp{5} \units{V/cm} \cdot (N_0/N)$ a typical time
$t_0 = 3\dexp{-12}\units{s}\cdot (N_0/N)$ and a typical density of
charge carriers $n_0 \equiv \epsilon_0E_0/\mathrm{e}l_0 
\approx 4.7\dexp{14}\units{cm^{-3}}\cdot (N_0/N)^2$, 
where $\mathrm{e}$ is the elementary charge, 
$N$ is the molecule number density of air and
$N_0$ is the number density at standard 
pressure and temperature, used
as an arbitrary reference.  We refer to \cite{Ebert2010/JGRA/c} for a physical
interpretation of these quantities and scaling laws.

Using these typical magnitudes one can build a dimensionless streamer model. 
We consider here a minimal model for air \cite{Ebert2006/PSST, 
Wormeester2010/arXiv} that also includes photo-ionization 
\cite{Luque2007/ApPhL}.  We are interested on the dynamics of
the streamer head, where impact ionization strongly dominates over
electron removal by dissociative attachment and the latter can be
safely neglected.  The governing equations are therefore
\begin{eqnarray}
\label{pde1}
\partial_t \sigma &=& D \laplacian \sigma + \mathbf{\nabla} \cdot (\sigma {\bf E})
+ S^{(ph)}+S^{(impact)}, \\
\label{pde2}
\partial_t \rho &=& S^{(ph)}+S^{(impact)}, \\
\label{pde3}
\laplacian \phi &=& \sigma-\rho,~~~{\bf E}= -\mathbf{\nabla} \phi.
\end{eqnarray}
Here $\sigma$ and $\rho$ are the (dimensionless) electron and ion
densities, $S^{(ph)}$ and $S^{(impact)}$ are, respectively, the
photo-ionization and impact ionization sources of electron-ion pairs,
$\phi$ is the electrostatic potential ${\bf
E}$ the electric field, and $D$ is a diffusion coefficient, 
taken as $D=0.1$.

There are several corrections to Townsend 
scaling \cite{Ebert2010/JGRA/c}.  One arises from collisional quenching 
of photo-ionization \cite{Liu2006/JPhD}, here included in $S^{(ph)}$, 
which contains an explicit dependence
on the air density.  A second correction, on which we focus here,
arises from the finite number of particles and the stochastic nature
of microscopic processes; this cannot be expressed in equations
(\ref{pde1})-(\ref{pde3}), because they only contain macroscopic quantities.
Rather we will take a
discretization of (\ref{pde1})-(\ref{pde3}) and convert it into a 
spatially extended stochastic model\cite{Gardiner2004/book}.

\mysection{Method} Since the importance of stochastic noise depends on the
particle number density our first step is to derive,
from the magnitudes described above, a
dimensionless parameter for the typical number of
charge carriers contained in a typical volume.  We define
$g \equiv n_0 l_0^3 = E_0 \varepsilon_0 l_0^2/\mathrm{e}$. 
This is the number of elementary charges that has to sit in each
area $l_0^2$ of an infinite charged plane to create a jump $E_0$ in
the electric field.   In air $g \approx 5700 \cdot (N_0/N)$; at an altitude
of $70\units{km}$, typical for sprites, $g \approx 10^{8}$.

The relative amplitude of the statistical fluctuations of a number 
$g$ of particles is $g^{-1/2}$.  The limit of negligible fluctuations
is $g^{-1/2} \ll 1$.  In air at atmospheric pressure, $g^{-1/2} \approx 0.01$ 
and stochastic noise is a relatively small correction on the fluid description.
However, as we will see, due to the strongly nonlinear nature of streamer
discharges, such small fluctuations can be amplified by strong
electric fields and alter significantly the propagation of a streamer.
For sprites, $g^{-1/2}$ is much smaller, about $10^{-4}$.

\begin{figure}
\includegraphics[width=\columnwidth]{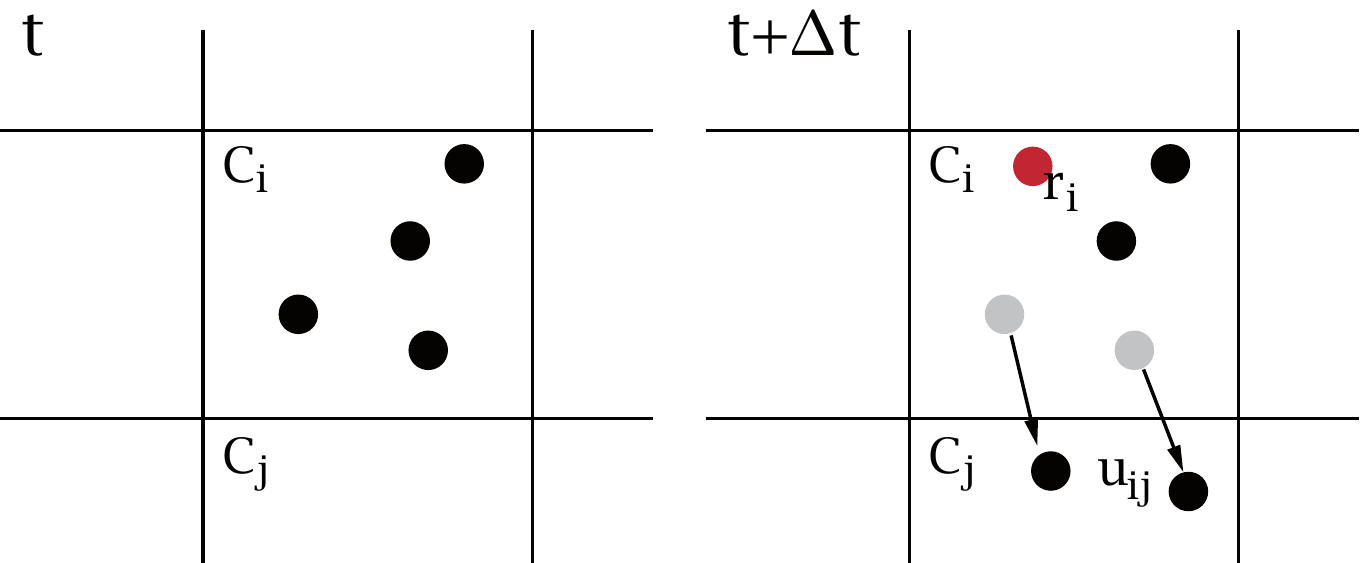}
\caption{\label{scheme} Scheme of an elementary time-step in the
  lattice model.  Each electron is represented here by a dot; note
  however that we do not keep track of every individual particle: only
  the number of particles in each cell.
  During the time-step from $t$ to $t + \Delta t$ we
  move $u_{ij}$ electrons from cell $C_i$ to $C_j$.  Meanwhile, 
  $r_i$ electrons arise in $C_i$ from impact ionization and
  photo-ionization.  }
\end{figure}

Now let us take a spatial discretization of (\ref{pde1})-(\ref{pde3}):
the simulation domain (see Fig.\ref{scheme}) is divided into cells 
$C_{i}$, each with a dimensionless volume 
$v_{i}=V_{i}/l_0^3$ ($V_i$ is the dimensional volume).
If we are given the dimensionless densities in each cell, 
$\sigma_i$ and $\rho_i$, we know, from our discretization, how to
calculate the left hand sides of (\ref{pde1})-(\ref{pde3}).  In
particular, we can calculate the source term 
$S_i = S^{(ph)}_i + S^{(impact)}_i$ and the flux from cell 
$C_{i}$ to each neighboring
cell $C_{j}$, which we denote $F_{ij}$.

But instead of densities one has a discrete
number of electrons and ions in each cell, $N^{(e)}_i$ and $N^{(i)}_i$; the
dimensionless densities are therefore 
$\sigma_i=N^{(e)}_i/n_0V_i = N^{(e)}_i/gv_i$, $\rho_i=N^{(i)}_i/gv_i$.
We may now use these to calculate the source terms and the fluxes.
The problem is now how to update the number of particles as time evolves.
To simplify the description, let us first describe a 
time-stepping that will tend to a forward Euler discretization with time
step $\Delta t$.  

We look first at the source terms.  Let $r_i$  be the number
of electron-ions pairs created in $C_i$ during a time $\Delta t$.
This quantity follows a Poisson 
distribution with average $\lambda_i=gv_iS_i\Delta t$.  So in the
simulation we draw for each $i$ a
sample from the Poisson probability distribution
$p(r_i) = \lambda_i^{r_i}e^{-\lambda_i}/r_i!$.

The transport terms are slightly more complicated because the
number of electrons must be conserved.  We interpret the
$gv_iF_{ij}\Delta t$ as the average number of electrons that flows from
$C_i$ to $C_j$ during $\Delta t$ (note that $F_{ij} \ne F_{ji}$).
Thus the probability for an electron in $C_i$ at $t$ to end in
$C_j$ at $t+\Delta t$ is $p_{ij}=N^{(e)}_i F_{ij}\Delta t / gv_i$ for $i \ne j$.
But since the electron must end somewhere, $p_{ii} = 1 - \sum_{j\ne i} p_{ij}$
\footnote{Note that nothing assures us that 
$p_{ii} > 0$, although it would be if $\Delta t \to 0$.  The solution that
we implemented in that case is to set $p_{ii} = 0$ and renormalize the rest of
the $p_{ij}$ such that $\sum_{j}p_{ij} = 1$}.
We can use these $p_{ij}$ to obtain the number of electrons moving
from $C_i$ to $C_j$, that we denote $u_{ij}$.  The
probability distribution for the electrons exiting $C_i$ is the multinomial 
distribution with $\sum_j u_{ij} = N^{(e)}_i$.

Now we have all the ingredients to update the particle numbers as
\begin{equation}
N^{(e)}_i(t + \Delta t) = N^{(e)}_i(t) + r_i + 
\sum_j \left(u_{ji} - u_{ij}\right)
\end{equation}

As mentioned above, this scheme tends to an explicit Euler time discretization
of the fluid equations as $g^{-1/2}\to 0$.  But we can also design a two-step
time updating that tends to second order Runge-Kutta
\cite{Montijn2006/JCoPh} if we (a) use the
particle numbers at $t$ to calculate $S_{i}(t)$ and $F_{ij}(t)$, (b) perform a 
half-step using $\Delta t/2$ to update the particle numbers, (c) use the
new particle numbers to obtain $S_{i}(t + \Delta t/2)$ and 
$S_{i}(t + \Delta t /2)$ (c) define 
$S'_{i}(t) = (S_{i}(t) + S_{i}(t + \Delta t))/2$, 
$F'_{ij}(t) = (F_{ij}(t) + F_{ij}(t + \Delta t))/2$ and use them to perform
a step $\Delta t$.  Note that in principle step (b) could also implement 
a standard continuum step, since nothing forces us to preserve an integer
number of particles in that intermediate step; we opted however to use the same
stochastic step in both stages of the algorithm.

We have not yet mentioned the spatial discretization to 
calculate $F_{ij}$.  The reason is that the scheme is flexible on
that.  We used here the scheme described in \cite{Montijn2006/JCoPh}.
This is a flux-limited, nonlinear discretization schemes 
and it poses an additional difficulty: it sometimes leads to 
negative $F_{ij}$ which cannot be interpreted as a 
probability.  In that case, we rearrange the fluxes by letting 
$F_{ij} \to F'_{ij}$, with $F'_{ij}=0$, 
$F'_{ji} = F_{ji} - F_{ij}$ until no negative fluxes remain.

\mysection{Results} We now use the adaptive grid
refinement and the fluxes and reaction terms from
\cite{Montijn2006/JCoPh} and the three-dimensional cylindrical mesh of 
Ref.~\cite{Luque2008/PhRvL}. 
As a first application, let us analyze the initiation of breakdown
in a small plane-to-plane geometry with a potential difference of
$16 \units{kV}$ between two electrodes separated by 2 mm of air.
The simulated volume is discretized into
cells $\Delta r = \Delta z = 8 \units{\mu m}$, $\Delta \theta = 2\pi /
64$.  As initial condition, we set a neutral 
hemispherical gaussian seed at $z = r = 0$ (positive electrode) 
containing $\sim 6\dexp{5}$ electrons.

\begin{figure}
\includegraphics[width=\columnwidth]{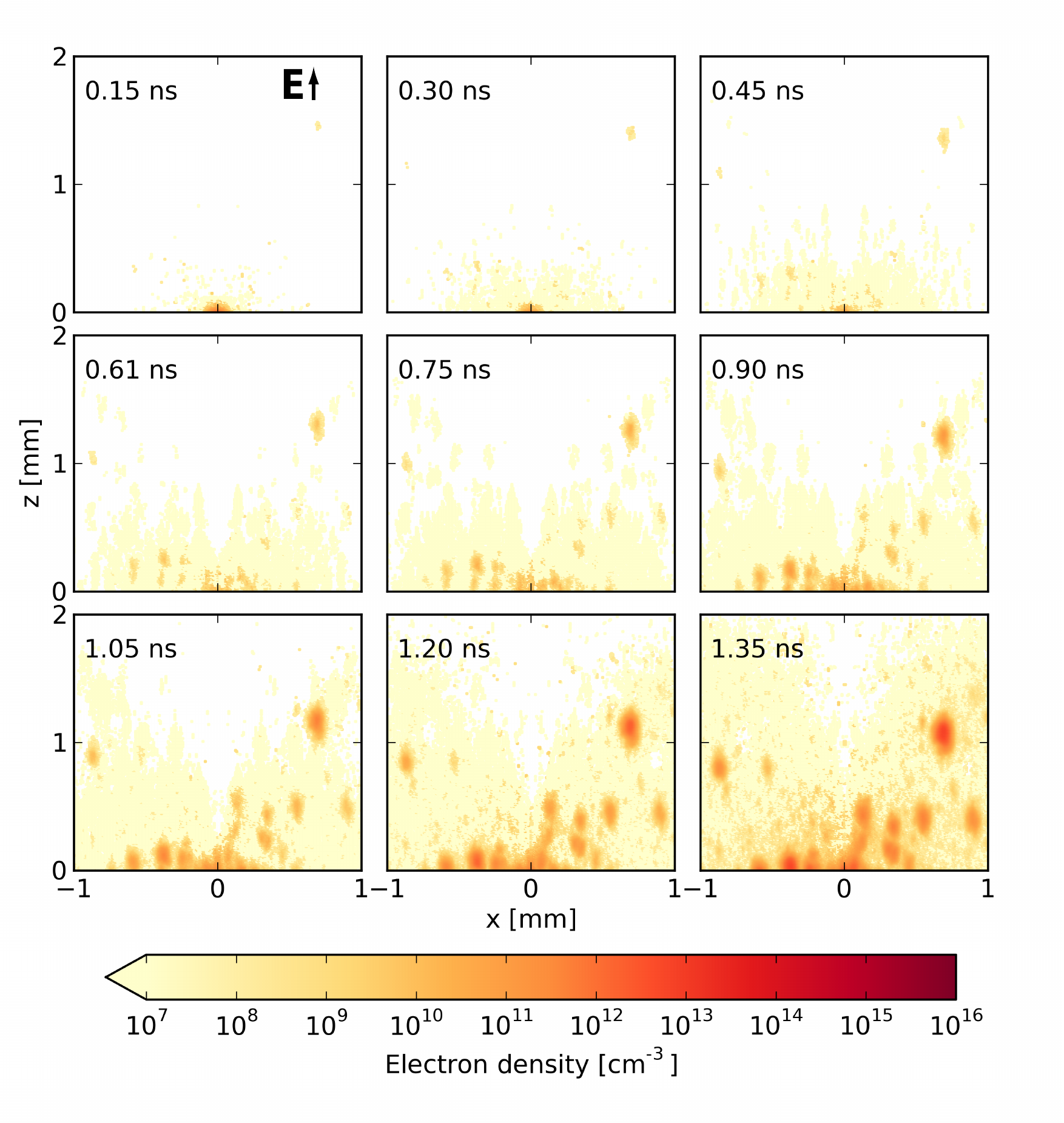}
\caption{\label{homogen} Snapshots of the cross section of the 
electron density in the $y = 0$ plane in a plane-plane, 2 mm gap with a
potential difference of 16 kV.  Multiple avalanches start out of 
seeds produced by photo-ionization; the electric breakdown extends to the
complete volume and no actual streamer is initiated. 
A movie of this simulation is available in the auxiliary material.
}
\end{figure}

Figure \ref{homogen} shows the evolution of a cross-section of the 
electron densities up to $1.35 \units{ns}$.  
In that short time-span, a multitude of avalanches
seeded by photo-ionization has developed.  A very similar evolution is
observed in the Monte Carlo simulations of \cite{Li2011/arXiv};
both results show that in strong electric fields noise weakens or even
prevents the formation of streamers.

\begin{figure}
\includegraphics[width=\columnwidth]{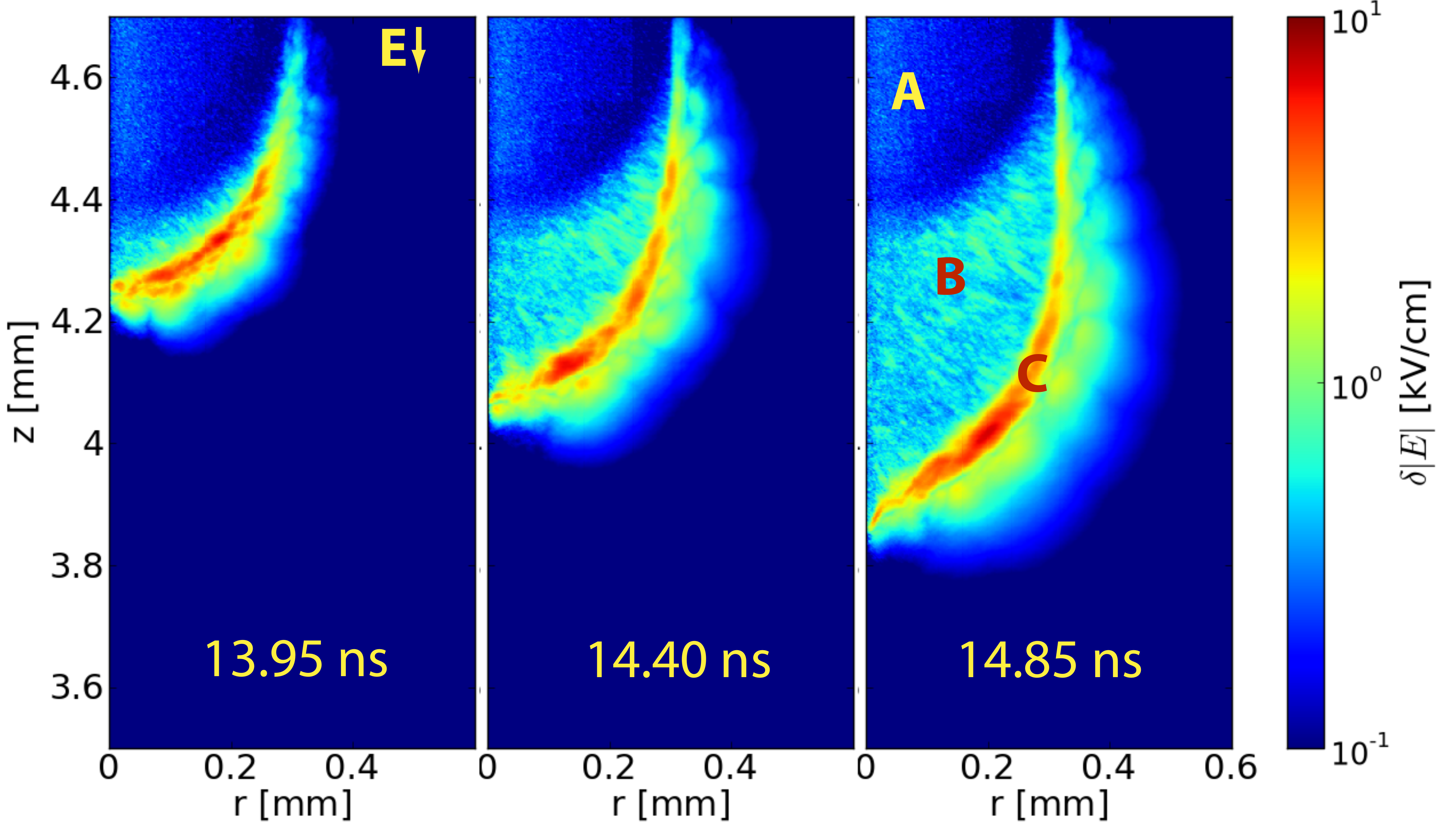}
\caption{\label{deltas} Highest deviation of the absolute value of the
electric field from its azimuthal average at three snapshots.  
The volume labelled A corresponds to the evolution of the streamer up
to $t = 13.5 \units{ns}$, with noise switched off.  When we switch on
noise, the streamer body (B) develops low-amplitude, small-scale
fluctuations.  The streamer surface close to the tip (C) has fluctuations 
with a higher amplitude and an auto-correlation length of about one tenth of a 
millimeter.  The amplitude grows but in this simulation it remains small
small compared with values of some hundreds of kV/cm in that area.}
\end{figure}
 
The situation changes however for a streamer in a point-plane
geometry.  Then only in a small volume is the
electric field high enough to produce significant ionization.

We run a simulation with point-plane electrodes implemented as in
\cite{Luque2008/JPhD}.  The needle is $2 \units{mm}$ long and its tip
is separated from the plate by $7.2 \units{mm}$.  As initial condition
we set a semi-spherical neutral gaussian ionization seed at the needle
tip with a radius of $73.6 \units{\mu m}$ and a peak ionization
density of $4.7\dexp{18} \units{cm^{-3}}$.
The needle has a
positive potential of $10.5 \units{kV}$ relative to the lower electrode.  
We used an adptive refinement strategy
\cite{Montijn2006/JCoPh} with coarsest grid 
$\Delta z = \Delta r = 40 \units{\mu m}$ and finest grid
$\Delta z = \Delta r = 2.5 \units{\mu m}$ 
\footnote{In the adaptive refinement algorithm one has to
  interpolate densities from coarser to finer grid.  We adapted this
  to our discrete algorithm by interpreting densities as probabilities
  and again sampling from a multinomial distribution.  Thus, all cells
  contain a discrete number of particles and this number is preserved
  across nested grids.}.
Since full 3d simulations are too demanding, we chose to
run the simulation with cylindrical symmetry and $g^{-1/2}=0$
up to $t = 13.5 \units{ns}$.  Then we remove the constraint of cylindrical
symmetry and introduce stochastic noise at the level expected at
atmospheric pressure ($g=5700$).  The continuous density at each cell is then
interpreted as an average and discrete numbers of particles 
are obtained by drawing random samples from a Poisson distribution
with this average.

To represent the evolution of this 3d simulation
that deviates only slightly from perfect cylindical symmetry,
 let us consider
the average of some quantity around the azimuthal angle,
$\langle u\rangle_\theta = \frac{1}{2\pi} \int_0^{2 \pi}u(r, z, \theta)d \theta$.
The deviation from symmetry can then be defined as 
$\delta u = \max_\theta (u - \langle u\rangle_\theta)$.
Figure~\ref{deltas} shows $\delta |E|$ at three instants of
time after noise is introduced into the simulation.

This simulation is highly demanding: the short streamer evolution
represented in Fig.~\ref{deltas} took about 5 weeks using two
dual-core 3 GHz AMD Opteron processors.  We
could not run the simulation long enough to observe
actual branching;  nevertheless, we can use the present result for a first
quantitative estimation of the time needed to branch.

\begin{figure} 
\includegraphics[width=\columnwidth]{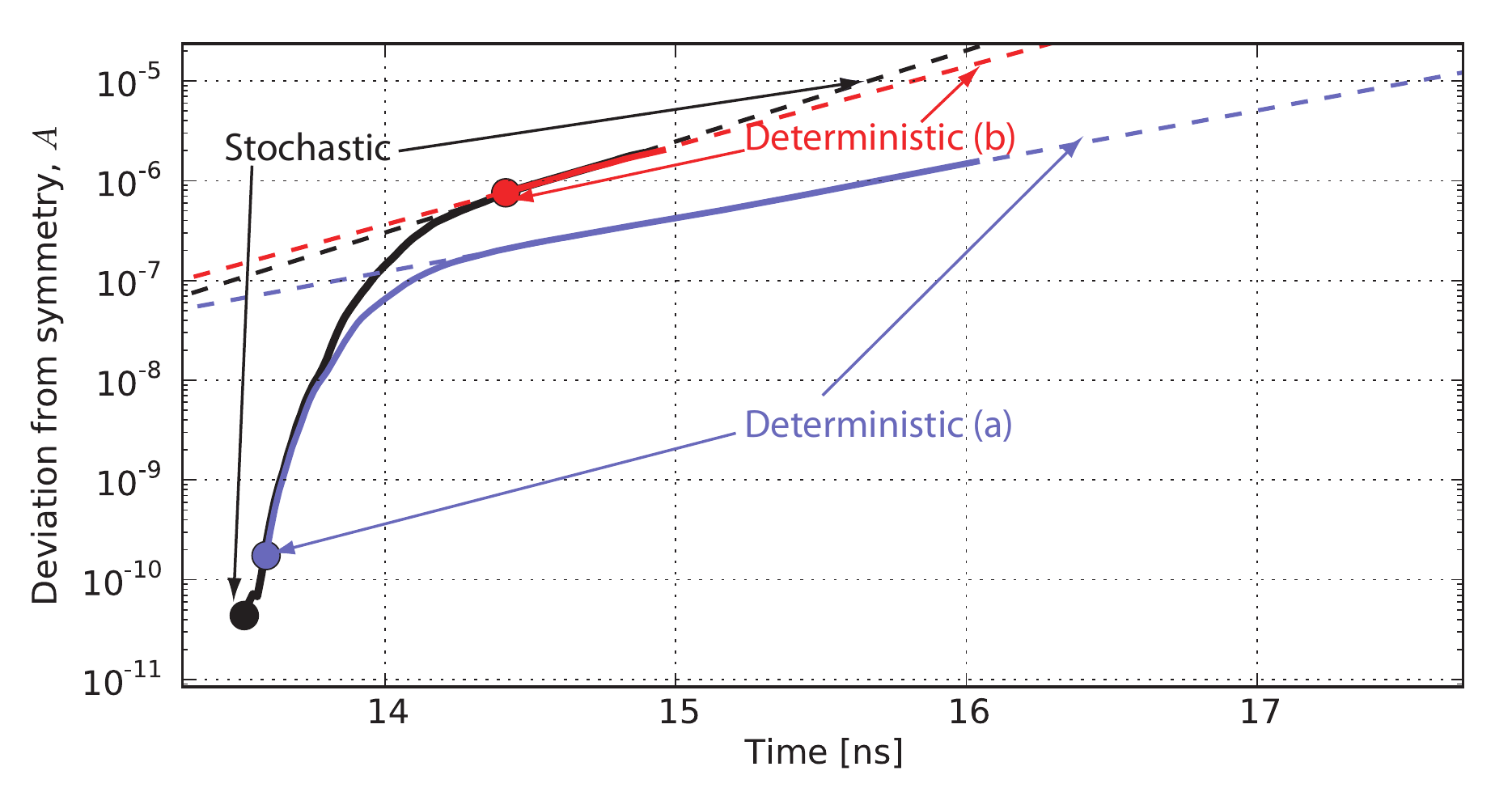} 
\caption{\label{afit} Evolution of $A$ as defined in the text.  
We show the evolution of three runs: in all runs the system was
deterministic and cylindrically symmetrical up to time 13.5 ns.  After this point we 
add stochasticity and allow deviations from  cylindrical symmetry.  
In the curves marked as deterministic, we remove noise again at the time 
marked with the dots while keeping the unsymmetric perturbations that
have evolved up to that time.  
Within
this limited timeframe, $A$ seems go through a transient, fast
growth phase until it settles into an exponential growth.  The dashed line
fits this second phase.  For the completely stochastic run the best fit is
$A = ae^{t/\tau}$ with $a = 5\dexp{-20}$, 
$\tau=0.48 \units{ns}$.  Extrapolating, this predicts branching at
$t \sim -\tau \log a \approx 21 \units{ns}$.}
\end{figure}
 
Let us look at the Fourier transform of the electron density $\sigma$ along the
azimuthal coordinate $\theta$, 
$\tilde{\sigma}(r, z, k) = 
\frac{1}{2\pi} \int_0^{2\pi} d\theta n_e(r, z, \theta) e^{-ik\theta}$.
We can define the ``total spectral content'' of mode 
$k$ of the electron density as 
$W_k = 2\pi \int_{-\infty}^{+\infty}dz\, \int_0^{\infty}dr\, r
\tilde{\sigma}(r, z, k)$.  If the streamer is close to cylindrical
symmetry (i.e. it is far from a branching state), 
$|W_0| \gg |W_k|$ for all $k > 0$.  Hence if we define
$A = \sum_{k>0} |W_k|^2 / |W_0|^2$ we can postulate that the
condition for branching is $A \sim 1$.

In Figure \ref{afit} the value of $A$ during the simulated time
frame is plotted as a black line.  During the full evolution we 
find $A \ll 1$, i.e. the streamer
was at all times far from branching.  However we
can use $A(t)$ to obtain a first estimation of the
branching time.  After a transient, $A$
growths exponentially; extrapolating this growth to $A \sim 1$ we
can roughly estimate 
the branching time as $t_{branch} \sim 21 \units{ns}$.  The
streamer velocity is
approximately $v = 0.32 \units{mm/ns}$ and hence after the
introduction of noise the streamer would
run for about $2.4 \units{mm}$ before branching.
The measurements in \cite{Briels2008/JPhD, Nijdam2008/ApPhL} give a
ratio of the branching distance to the streamer diameter of about
12-15.  This is relative to the radiative diameter, estimated to be
about half of the electrodynamic diameter
\cite{Pancheshnyi2005/PhRvE}; with that estimation, our value is about
8.   One must also take into account that (a) short
branching distances are harder to measure and therefore the average in
\cite{Briels2008/JPhD, Nijdam2008/ApPhL} may be slightly overestimated 
and (b) both in our model and in observations branching distance is random:
we are comparing the result of only one simulation with an average
over many measurements so a certain discrepancy seems natural.

However, our estimation of a branching distance is close to the
measured value, suggesting
that noise resulting from the finite number of particles is relevant
in streamer discharges.  To measure the relevance
of a persisting noise compared with the inherent growth of cylindrically 
perturbed modes we performed two simulations in which noise is removed after
some time.  These are shown in the two curves of Figure \ref{afit} labelled 
as deterministic (a) and (b).  We see that noise always increases the
growth rate of the deviations but also that even a relatively small
amount of noise during a short time (a) is enough to trigger an
instability that would eventually lead to streamer branching.
Longer simulations must be performed to check that the evolution suggested in 
Fig.~\ref{afit} continues until the streamer branches.
 
Note that at lower pressure, such as those in the high
layers of the atmosphere where sprite streamers are commonly observed,
stochastic noise is weaker ($g^{-1/2} \approx 10^{-4}$).  Although
the ratio between diameter and branching distance has not yet been
measured in sprites, they also branch frequently \cite{McHarg2010/JGRA/c}.
However, in the upper atmosphere, other sources of stochasticity may
be relevant, such as cosmic rays and atmospheric inhomogeneities.

\begin{acknowledgments}
\emph{Acknowledgments:} This work was supported by the Spanish Ministry of Science and
Innovation, MICINN under project AYA2009-14027-C05-02 and 
and by the Junta de Andalucía, Proyecto de Excelencia FQM-5965.
\end{acknowledgments}

\newcommand{\jgr}{J. Geophys. Res.}
\newcommand{\grl}{Geophys. Res. Lett.}
\bibstyle{aip}
\bibliography{Everything}

\end{document}